# Fundamental lower size limit in wavelength selecting structures[*]

A. Driessen, H.J.W.M. Hoekstra, D.J.W. Klunder and F.S. Tan
*MESA+ Research Institute, University of Twente, P.O. Box 217, 7500 AE Enschede, The Netherlands*

## Abstract

The fundamental lower size limit in wavelength selecting structures is explored with the aid of the Heisenberg uncertainty principle. The analysis shows that for a given wavelength selectivity resonating structures with optical feedback have the smallest dimensions. Experimental results obtained with integrated optics microring resonators (Q ~ 3.4 x $10^4$) confirm the analysis. In addition, a discussion is given on the validity of the uncertainty principle in terms of hidden variables or restricted knowledge on the system in question.

---

[*] This is the first of a collection of four studies dealing with the weird properties of a photon, compiled in 2017 by Alfred Driessen. The original version of 2003 has not yet been published and the current version is essentially unchanged. No update of the references is given.

The collection includes:

1. Fundamental lower size limit in wavelength selecting structures, A. Driessen, H.J.W.M. Hoekstra, D.J.W. Klunder and F.S. Tan, 2003 (unpublished)

2. Evidence for nonlocality and nontemporality of a single photon, A. Driessen, 2003 (unpublished)

3. Propagation of short lightpulses in microring resonators: Ballistic transport versus interference in the frequency domain, A. Driessen, D.H. Geuzebroek, E.J. Klein, R. Dekker, R. Stoffer, C. Bornholdt Optics Communications 270 (2007) 217-224. doi:10.1016/j.optcom.2006.09.034
https://www.researchgate.net/publication/221661942_Propagation_of_short_lightpulses_in_microring_resonators_Ballistic_transport_versus_interference_in_the_frequency_domain

4. The strange properties of the photon: a case study with philosophical implications, Alfred Driessen, Talk presented at Journée « Science, raison et foi » on Causalite, Temps et Origine de l'Univers, CASTELVIEIL, MARSEILLE, November 12 , 2013,
https://www.slideshare.net/ADriessen/the-strange-photon

## 1. Introduction

The Heisenberg uncertainty principle (HUP) is still subject of fundamental discussions in the field of quantum mechanics. There are principally two views. For the majority the principle is accepted as a law of nature arising as a consequence of the statistical character of quantum mechanics, whereas for others it is an expression of the insufficient knowledge of the system under study. In their opinion hidden variables or some new physical insight could circumvent this deficiency. In this fundamental aspect the discussion is closely related to the EPR paradox [1] and the contributions of Bell [2] and others.

In the following the argumentation in favor of the general validity of the HUP is based on an indirect approach: assuming its fundamental validity, rules for solving a physical problem are derived and experimentally verified. Exploiting the uncertainty in a variable, say time, the conjugate variable, energy, can be determined with great accuracy. If the uncertainty would only be an uncertainty with respect to the observer or the experimentalist, changing this subjective uncertainty would not change the outcome of the experiment for the conjugate variable. If it does, however, another evidence for the Heisenberg principle as being a fundamental law of nature is given.

The following considerations have not only a pure academic motivation; they are also related to an important technological problem in the field of Very Large Scale Integrated (VLSI) photonics. Analogue to VLSI electronics where millions of electronic elements, e.g. transistors, are integrated on a few cm$^2$, guided wave optics has meanwhile gained certain maturity. The number of optical functions are still relatively small, several hundreds at best, but new approaches like high-index contrast waveguides – photonic wires – and photonic crystals open the route to higher integration. A key function in these circuits is wavelength selection that allows combining and selecting a large number of wavelengths in dense Wavelength Division Multiplexing (WDM) networks. A recent example of such a WDM device is given by Takada et al. [3], who realized a 1000 channel de-multiplexer with a channel spacing of 0.2 nm corresponding to 25 GHz. The relative large chip-area - a complete 4 inch wafer is needed - excludes, however, applications as a low-cost device in VLSI photonics. The size, or specifically the chip area needed for a single wavelength-selecting element, is one of the critical parameters for VLSI photonics.

For the optimum solution of this technical problem it is necessary to address the fundamental physical principle responsible for wavelength selection. Once this is known it will be possible to design structures that make an as efficient as possible use of this principle. The relevant fundamental principle is the HUP applied to photons [4], as it relates the uncertainty or spread in wavelength to a principle length uncertainty in the optical path, as will be shown below. The application of the HUP to optical set-ups has been addressed by various authors, among others recently by Stelzer and Grill [5] who estimated focal spot dimensions. In the following we analyze the fundamental physical lower size limits of wavelength selecting devices by applying the HUP and demonstrate that integrated optics microresonators (MRs) as realized for example by us in the silicon oxynitride technology [6] approach this fundamental limit.

## 2. The Heisenberg uncertainty principle applied to wavelength selecting devices

In the analysis of the physics of wavelength selecting devices one may consider light as an electromagnetic wave that propagates according to Maxwell's equations. This approach is particularly suited for numerical calculations and design of waveguiding structures that take into account the detailed geometry and the materials parameters. The particle picture, i.e. considering photons and entering the field of quantum mechanics, however, can give more insight in the physics [7]. Therefore we will use this approach in an elementary way with just sufficient parameters to understand the fundamental limits of this class of devices.

In the quantum mechanical picture light can be considered as particles, whose energy **E** is related to the angular frequency $\omega$ by:

$$\mathbf{E} = \hbar \omega \qquad (1)$$

with $\hbar = h/2\pi$ and $h$ the Planck constant. If we assume the fundamental validity of the HUP principle for photons one can relate the uncertainty in energy to the uncertainty in a critical time or duration **t** by [8]:

$$\Delta \mathbf{E}\, \Delta \mathbf{t} \geq \hbar/2 \qquad (2)$$

Applying Eq. (1) to a photon and assuming a minimal uncertainty one gets:

$$\Delta \omega\, \Delta \mathbf{t} = 1/2 \qquad (3)$$

The angular frequency $\omega$ is related to the wavelength $\lambda_0$ and the speed of light *c*, both *in vacuo* by:

$$\omega = 2\pi c / \lambda_0 \qquad (4)$$

in addition, the uncertainty in time can be related to an uncertainty in length:

$$\Delta l = c \Delta \mathbf{t} \qquad (5)$$

where $\Delta l$ is a fundamental uncertainty in the optical path length of the photon. Coming back to our problem we see with Eq. (3) that the fundamental limit in the uncertainty in wavelength $\Delta \lambda_0$ is related to:

$$\Delta \lambda_0 = \lambda_0^2 / (4\pi c\, \Delta \mathbf{t}) \qquad (6)$$

This relation - essentially equivalent to the HUP - states that by maximization of the uncertainty in time one obtains the smallest uncertainty in wavelength at a given wavelength. We conclude therefore that assuming the fundamental validity of the HUP the fundamental physical principle of wavelength selecting structures is the introduction of a fundamental time uncertainty in the lightpath or alternatively, with Eq. (5) a fundamental uncertainty in the optical path length.

Born and Wolf [9] derived Eqs. (3), (5) and (6) directly without using the HUP. In that case they interpret $\Delta t$ and $\Delta l$ as the coherence time and coherence length of the lightsource. When analyzing the resolving power of a grating, Born and Wolf state: *The resolving power is equal to the number of wavelengths in the path difference between the rays that are diffracted*. They confirm, at least for a grating, that the principle uncertainty in the optical path length is a measure for the resolving power. A simple geometrical analysis shows that with Eq. (6) – that was based on the Heisenberg uncertainty principle - the same expression for the resolving power of a grating can be derived as given by Born and Wolf.

In practice, arrayed waveguides gratings (see e.g. [4]) are often used as wavelength selecting structures in WDM systems. In these structures light passes through an array of waveguides with carefully designed differences in optical path length. Also here the principle uncertainty in the optical path length defines the resolving power. For it is principally undetermined in which of the many parallel waveguides - each having a slightly different path length - the photon is propagating. In the case of the high performance device by Takada et al [3] the structure is quite large, about 50 cm$^2$. When increasing the index contrast, structures as small as 0.7 mm$^2$ have been realized for arrayed waveguides filters [10].

Now coming back to the problem of the reduction of size for wavelength selective devices it is obvious that optical resonators where photons pass several times a certain light path are favored structures for obtaining large $\Delta t$. The simplest form, the Fabry-Perot (FP) resonator, acts as a periodic wavelength filter, where the wavelength selectivity, the full width at half maximum (FWHM) of the resonance lines is inversely proportional to the finesse *F*. In a FP the principal time uncertainty arises from the principal uncertainty in the number of times the photon is oscillating forth and back before leaving the resonator. Only known is the average number *m*, which is closely related to the finesse: $m=F/2\pi$. The time uncertainty therefore is given by:

$$\Delta t = F\,n\,d\,/\,\pi\,c \qquad (7)$$

where *n* is the index of refraction and *d* the cavity length. With Eq. (6) one sees immediately that the uncertainty in wavelength has the expected reverse proportionality with regard to *F*.

As the proper working of a FP assumes parallel light a reduction in beam diameter in order to obtain a reduction in size of the FP will eventually encounter a limit when the beam diameter approaches the wavelength. In that case the concept of geometrical optics does not hold and diffraction will cause curved wavefronts and loss of parallelism. Singlemode waveguiding structures, where light is confined by total internal reflection or - in the case of photonic crystals - by Bragg reflection and propagates nearly loss-less, overcome the diffraction problem. Guided wave optical resonators therefore are best candidates for restricted-size wavelength selecting devices.

## 3. Experiments with optical microresonators

For the technological realization of guided wave microresonators the FP puts severe demands on the optical quality of the reflectors. In practice it is easier to realize high finesse spherical or cylindrical resonators where light propagates in whispering gallery modes[11]. Spherical microresonators in the form of liquid or solidified droplets exhibit extreme high values for the quality factor Q or *F* as whispering gallery modes can propagate nearly loss-less at curved surfaces obtained by surface tension (see, e.g. Laine et al.[12] who measured a Q of $0.5 \times 10^8$ and a resonator linewidth of 5 MHz, or also Lin and Campillo[13]). In spite of promising attempts[14], the controlled incorporation of the micro spheres in a mass produced optical device seems not to be a realistic option. Planar cylindrical- or ring resonator devices[15] which can be realized by standard lithography are technologically much more promising and have due to the restriction to two dimensions a significantly smaller resonator volume. The circular ring geometry is preferred because circles have the largest minimum radius at a given device area and because the main source of optical losses are due to the small bending radius. Recently Hammer[16] proposed rectangular planar resonators, but up to now no high Q resonators have been demonstrated experimentally.

In the following, experimental results on integrated optics microresonators are presented that exhibit extreme high wavelength resolution [[17]]. In addition it is shown that the experimental wavelength selectivity is essentially equal to the fundamental limit set by the HUP as given in Eq. (6). The geometry of the ringresonator is depicted in Fig. 1. Light is guided to the resonator by the upper single mode waveguide that is weakly coupled through a 750 nm wide gap with coupling constant κ. After coupling to the resonator, light propagates in a whispering gallery mode with a certain roundtrip loss RTL determined by the intrinsic materials loss, the scattering loss, the bending loss through radiation and loss due to coupling to the symmetric input and output waveguides. By changing the wavelength of the incoming light one obtains Fabry-Perot-like spectra which clearly show the resonances with a FWHM linewidth Δλ separated by the free spectral range (FSR). An analytical model is developed that on the basis of the geometry, materials constants and the parameters determining the RTL is able to calculate the spectral response including the FSR, Q and the finesse *F* defined by *F* = FSR/*Δλ*.

Devices with a radius of 25 μm have been realized in silicon-nitride on a silicon wafer with a silicon-oxide buffer layer made be steam oxidization. The nitride has been deposited by low pressure chemical vapor deposition. The lateral structuring has been made by standard e-beam mask lithography and reactive ion etching. After fabrication the wafer has been cleaved vertical to the input/output waveguides to obtain good quality facets for coupling light in and out by microscope objectives. Spectra have been obtained by the aid of a narrow bandwidth tunable semiconductor laser operating around the for telecommunication important wavelength window at 1550 nm.

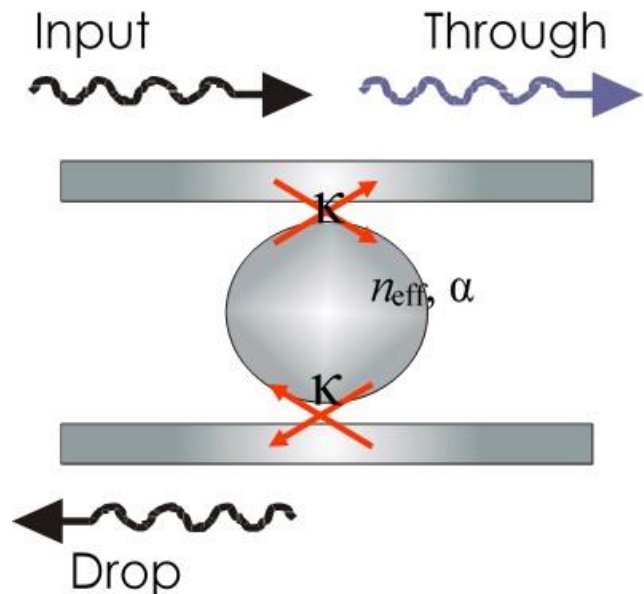

Fig. 1 Schematic view of a microring resonator, showing the port waveguides, the two coupling sections with coupling constant $\kappa$ and the resonator with a effective index $n_{eff}$ and absorption $\alpha$.

Fig. 2.a shows the transmission spectrum at the through port where clearly the sharp resonance peaks can be seen spaced by the FSR of approximately 8 nm. Fig. 2.b gives an enlargement around the resonance at $\lambda_0 = 1510$ nm with a finesse around 180 (Q ~ 3.4 x $10^4$) and a wavelength resolution of 44 pm. This is to our knowledge the highest resolution obtained in a planar device as small as 0.003 mm$^2$.

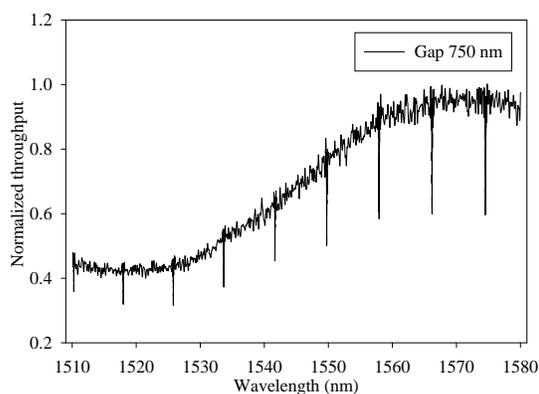
Fig. 2.a. Experimental transmission spectrum obtained a the through port of our MR with a radius of 25 $\mu m$.

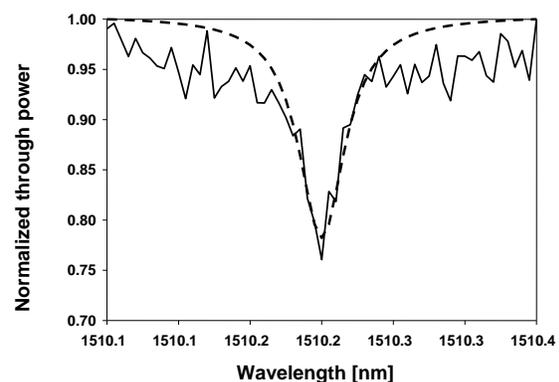
Fig. 2.b. Detail of experimental transmission spectrum obtained at the through port; solid line: experimental data; dashed line: fit obtained with analytical model resulting in a linewidth of 44 pm.

With the foregoing has been shown that the HUP has been a good guide to obtain compact wavelength selecting structures. It remains to be shown that the resolving power of a MR:

$$\Delta \lambda_0 = FSR / F \qquad (8)$$

leads to the formulation of the HUP as given in Eq. (6). With $FSR = \lambda_0^2 /(2\pi R n_{eff})$, where R is the radius of the MR and $n_{eff}$ the effective index of the resonator mode, and expressing the finesse in terms of the average number of roundtrips of the photon $m$ Eq. (8) can be written:

$$\Delta \lambda_0 = \lambda_0^2 /(2\pi R \cdot 2\pi m n_{eff}) \qquad (9)$$

For the uncertainty in optical pathlength one has: $\Delta l = m \cdot 2\pi R n_{eff}$, and with Eq. (5) one gets:

$$\Delta \lambda_0 = \lambda_0^2 /(2\pi c \Delta \mathbf{t}), \qquad (10)$$

which is within a factor of 2 identical to Eq. (6), an expression directly derived by the HUP. That means that the resolving power of a MR can in fact be derived by the HUP, if one neglects the missing factor of 2 probably caused by a non-consistent use of definitions in literature.

**4. Discussion**

We have shown that the HUP can guide the search for reduction in the size of wavelength selective structures. Resonating structures with optical feedback allow for the highest wavelength resolution at a given device volume, or in the case of integrated optics, at a given chip area. Our approach based on integrated optics MRs results in devices which perform in accordance with the fundamental lower size limit set by the HUP. Presenting explicit verification on only two possible wavelength selecting set-ups, gratings and MRs, we feel, after a rapid analysis, confident that also all other set-ups are performing according to the HUP.

With regard to the technical point of view we conclude that the compactness and resulting small chip area, about 0.003 mm$^2$, makes MRs promising building blocks of complex waveguiding structures needed for next generation optical networks.

In the introduction the fundamental discussion on the HUP has been mentioned. Is the uncertainty a consequence of a fundamental law of nature or just a lack of knowledge? Suppose the HUP is due to lack of knowledge of the system. In that case, if one knew the hidden variables of the system, the HUP could be circumvented. One then could know in principle energy and time of a photon (see Eq. (2)) with an arbitrary accuracy simultaneously. The uncertainty in the conjugate variables in the HUP therefore would not show the simple correlation in reality, only in the formulation of our restricted theory which does not include the hidden variables. If one changes experimentally one of the variables, say the uncertainty in time, one would not expect the change in the uncertainty in energy or wavelength as predicted by the HUP. Observing, however, experimentally that the correlation between the conjugate variables in wavelength selecting devices is according to the HUP (see the previous conclusion) the role of hidden variable is quite restricted as it adds nothing to the prediction of the HUP.

In summary one can conclude that the HUP is an efficient guide to conceive wavelength selecting devices and that consequently it is less probable that new insight will lead to physical laws in direct contradiction with the HUP. One may in addition exclude a prominent role for the observer or the status of his knowledge. For the wavelength selective devices based on the validity of the HUP perform independently of the observer. Hidden variable approaches as a deeper explanation of the HUP become therefore less convincing.

## 5. Acknowledgments